# Optical information transfer through random unknown diffusers using electronic encoding and diffractive decoding

Yuhang Li,[a,b,c] Tianyi Gan,[a,c] Bijie Bai,[a,b,c] Çağatay Işıl,[a,b,c] Mona Jarrahi,[a,c] Aydogan Ozcan[a,b,c,*]

[a]Electrical and Computer Engineering Department, University of California, Los Angeles, USA, 90095
[b]Bioengineering Department, University of California, Los Angeles, USA, 90095,
[c]California NanoSystems Institute (CNSI), University of California, Los Angeles, USA, 90095

**Abstract**. Free-space optical information transfer through diffusive media is critical in many applications, such as biomedical devices and optical communication, but remains challenging due to random, unknown perturbations in the optical path. In this work, we demonstrate an optical diffractive decoder with electronic encoding to accurately transfer the optical information of interest, corresponding to, e.g., any arbitrary input object or message, through unknown random phase diffusers along the optical path. This hybrid electronic-optical model, trained using supervised learning, comprises a convolutional neural network (CNN) based electronic encoder and successive passive diffractive layers that are jointly optimized. After their joint training using deep learning, our hybrid model can transfer optical information through unknown phase diffusers, demonstrating generalization to new random diffusers never seen before. The resulting electronic-encoder and the optical-decoder model were experimentally validated using a 3D-printed diffractive network that axially spans <70$\lambda$, where $\lambda = 0.75$mm is the illumination wavelength in the terahertz spectrum, carrying the desired optical information through random unknown diffusers. The presented framework can be physically scaled to operate at different parts of the electromagnetic spectrum, without retraining its components, and would offer low-power and compact solutions for optical information transfer in free space through unknown random diffusive media.

*Correspondence: Prof. Aydogan Ozcan  E-mail: ozcan@ucla.edu

## 1 Introduction

Free space optical communication has been an active research area and has gained significant interest due to its unique advantages, such as large bandwidth and high transmission capacity[1–4]. It has various applications in e.g., remote sensing, underwater communication, and medical devices[5–8]. One of the challenges that limit optical data transmission in free space with high fidelity is the existence of diffusive random media in the optical path, distorting the optical wavefront, which causes inevitable information loss during the wave propagation[9–11]. One approach to mitigate the negative impact of such random diffusers along the optical path has been single-pixel detection with several amplitude-only 2D encoding patterns used to illuminate the object/scene[12–16]. However, this approach is relatively slow since a series of amplitude-only illumination patterns need to be sequentially generated and transmitted as optical information carriers for each input object or message; furthermore, a digital reconstruction algorithm is required to reveal the input objects through a sequence of structured illumination patterns. Another potential solution involves using adaptive optics to correct distortions[17–19]; however, the spatial light modulators and feedback algorithms employed in adaptive optics increase the cost and complexity of such systems. Motivated by the widespread use of deep learning, recent research has also adopted deep neural networks in optical image transmission through scattering media[20–24], increasing the processing speed and robustness to the distortions caused by random diffusers. Nevertheless, all these methods need digital computation to decode, reconstruct or extract useful information from the distorted transmitted signals, which can be energy-intensive and time-consuming[25–28].

In this work, we demonstrate a jointly optimized electronic encoder neural network and an all-optical diffractive decoder model for optical information transfer through random unknown diffusers. This electronic-optical design (Fig. 1) is composed of a convolutional neural network (CNN) that encodes the input image information of interest (to be transmitted) into a 2D phase pattern, like an encrypted code, which is all-optically decoded by a jointly trained/optimized diffractive processor that reconstructs the image of the input information at its output plane, despite the presence of random unknown phase diffusers that are constantly changing/evolving (see Video 1). This all-optical diffractive information decoder consists of passive diffractive layers with a compact axial span of $<70\lambda$, and can rapidly decode the encoded optical information (corresponding to any arbitrary input object or message) through random phase diffusers without any digital post-processing. We experimentally validated the success of this approach using terahertz radiation ($\lambda = 0.75\,\mathrm{mm}$) and a 3D-printed diffractive network, demonstrating the feasibility of the diffractive decoder with electronic encoding for optical information transmission through random unknown diffusers. The optical decoder of the presented hybrid model can be scaled physically to operate at different parts of the electromagnetic spectrum without the need for retraining its diffractive features, and might find various applications in biomedical imaging and optical communications through random diffusive media.

## 2 Results

*2.1 Design of a diffractive decoder with electronic encoding for optical information transfer through unknown random diffusers*

Fig. 1(b) depicts the operational pipeline of the presented electronic-encoder and optical-decoder model for optical information transfer through random phase diffusers. A CNN-based encoder (see the Materials and methods section) was trained to convert any given input image to be transferred into a phase-encoded pattern, illuminated by a uniform plane wave. The corresponding optical field distorted by random unknown phase diffusers was then decoded by the diffractive network to all-optically recover the original image at its output field of view (FOV).

First, we analyze the impact of the encoder CNN on the optical information transfer through unknown random diffusers present in the optical path, and quantitatively explore its necessity, as opposed to a diffractive decoder that is trained *alone*. For this analysis, we compared it against the architecture of our previous work[29,30], which was used to see amplitude objects through random diffusers using a diffractive neural network, as shown in Fig. 2(a). Without any encoder neural network present, this diffractive processor could see through random unknown diffusers after its training with hundreds to thousands of examples of random diffusers, successfully generalizing to see through new random diffusers never seen before. However, with the increase in the distance between the input objects and the random phase diffuser plane, a performance drop was observed, as shown in Fig. 2(c). Here we used the Pearson Correlation Coefficient (PCC) to assess the quality of the output images synthesized by the diffractive network as a function of the Fresnel number ($F = \frac{a^2}{d\lambda}$), where $a$ is the half-width of the input object field-of-view and $d$ is the axial distance between the input plane and the unknown phase diffuser plane, which was swept from $d \sim 13\lambda$ to $93\lambda$ (see the Materials and methods section). These comparative results shown in Fig. 2(a, c) reveal that, while a diffractive decoder alone can be trained to successfully generalize to see through unknown random diffusers, its all-optical image reconstruction quality drops as $d$ increases to more than ~50 $\lambda$ (corresponding to $F < \sim 20$) for a diffuser correlation length of $L \sim 5\ \lambda$. For example, consider the extreme case of $d$ approaching 0 (i.e., $F \to \infty$): in this case, it is expected that a diffractive decoder alone can create a diffraction-limited image of the object at its output FOV since the problem of this special case boils down to a spatially-incoherent[31] imaging system where the impact of a random phase diffuser rapidly changing at $d = 0$ can be considered as spatially incoherent illumination of a static amplitude only object. As $d$ increases, however, each randomly changing phase grain at the diffuser plane will start introducing perturbations to spherical waves that originate from a larger number of the object points at the input plane due to diffraction across $d$, making the image reconstruction a more difficult problem. This is at the heart of the performance reduction observed in Fig. 2(a, c) for the diffractive decoder alone operating at $d > \sim 50\ \lambda$ ($F < \sim 20$) for $L \sim 5\ \lambda$. The all-optical image reconstruction examples of the handwritten digit '8' with only the diffractive decoder layers (no electronic encoder network) also confirm this conclusion in Fig. 2(a), where the contour of the digit '8' became unrecognizable when $F < \sim 20$.

To break through this limitation and extend the capabilities of diffractive neural networks to communicate through random unknown diffusers, we jointly trained an electronic encoding neural network to collaborate with the all-optical diffractive decoder network for transferring optical information through random unknown phase diffusers, covering a much larger span of $d$ and $F$ as shown in Fig. 2(b, c). To achieve this enhanced performance with the hybrid (electronic-optical) model, a data-driven supervised learning strategy was utilized to transfer any optical information or message of interest through random unknown phase diffusers. Specifically, we trained a CNN-based electronic neural network (see Supplementary Fig. S1), which encoded the input samples from the MNIST dataset into $[0 - \pi]$ phase-only 2D patterns (codes), and a 4-layer diffractive network was jointly optimized with this CNN to be able to decode the optical fields distorted by random unknown diffusers at the output field-of-view. This decoding part and the reconstruction of the object images are performed all-optically and rapidly completed at the speed of light propagation through thin passive layers, without any external power source. Multiple ($n = 20$) randomly generated diffusers, modeled as thin phase masks, were introduced in each training epoch to build generalization capability for both the electronic encoder and the diffractive decoder against the distortions caused by random phase diffusers, as shown in Fig. 3(a). The correlation length ($L$) was used to characterize random diffusers with respect to their effective grain size, and all the random diffusers used in the training and blind testing were set to have the same correlation length, $L=5\lambda$ (see the Materials and methods section). During the training stage, handwritten digits were fed to the electronic encoder, and the resulting encoded phase patterns were illuminated by a uniform plane wave, propagating through the corresponding random diffuser and the successive diffractive layers to form the intensity profiles at the output plane. The height values of the diffractive features at each layer were adjusted through stochastic gradient descent-based error backpropagation by minimizing a PCC-based loss function between the target optical information and the decoded output intensity profiles (see the Materials and methods section). The encoder and the decoder were trained simultaneously for 100 epochs, i.e., $N = 100n = 2,000$ different random diffusers were used during the entire training process. Figure 3(b) shows the resulting diffractive layers of the trained diffractive decoder.

To demonstrate the efficacy of this trained hybrid electronic-encoder and optical-decoder for transferring optical information through random phase diffusers, we first compared its performance to that of a conventional lens-based image transmission system (see the Materials and methods section). The imaging results of the same objects through the same diffusers, as shown in Fig. 3(d), exhibited blurry intensity profiles of handwritten digits '2', '7', and '8' when captured by an ideal lens. These blurry images clearly illustrate the negative impact of the random phase diffusers within the optical path, making it impossible to recognize the transmitted images with the naked eye. In contrast, the jointly trained electronic-optical system, comprising an electronic encoder/front-end and an optical diffractive decoder/back-end, can successfully transmit and receive these images through unknown new phase diffusers, as depicted in Fig. 3(c), indicating the strong resilience of the hybrid system against the distortions caused by random diffusers during the signal transmission process.

To shed more light on the generalization ability of the hybrid model, we further tested it with additional handwritten digits sampled from the test set that were never used during the training stage; these test objects were individually distorted by 2 newly generated diffusers that were never encountered during the training (termed as *new* diffusers), as shown in Fig. 4 left two columns. Additional results of this hybrid electronic-optical model for transferring new optical images of interest through random unknown phase diffusers that are constantly changing are shown in Video 1. All of these resulting output images are easily recognizable compared to the blurry counterparts of the ideal lens, revealing the generalization of the trained electronic-encoder and optical-decoder model to new objects and new random phase diffusers that were never seen before.

To provide examples of "external generalization" to test objects of different types, we used binary gratings with $8\lambda$ linewidth, which were not included in the training dataset; such linear gratings are vastly different compared to handwritten digits used during training and were employed here to exemplify the external generalization of the jointly trained system for transferring spatial information corresponding to different types of objects through *new* random diffusers (see Fig. 4 right two columns). After being encoded by the electronic encoder CNN and transmitted through the random phase diffusers, the gratings were successfully decoded by the diffractive optical decoder, even though our electronic-optical model was trained using only MNIST handwritten digits. The successful transmission of different types of object information through random, unknown phase diffusers indicates that our hybrid electronic encoder and optical decoder system jointly approximates a general-purpose optical information transfer system, without overfitting to a special class of objects. The spatial resolution of this jointly trained electronic-optical model can be further improved by including training images that contain higher spatial frequency information[32] as opposed to using relatively low-resolution training image sets such as MNIST.

*2.2 Impact of the number of diffractive layers on the optical information transfer fidelity*

The depth advantages that deeper diffractive architectures possess include better generalization capacity for all-optical inference tasks, which has been supported in the literature by both theoretical and empirical evidence[29,31,33–37]. To quantitatively evaluate the impact of the number ($K$) of diffractive layers on the accuracy of optical information transfer through unknown random diffusers, we trained three hybrid models, where the architectures of the electronic encoder were kept identical, but the diffractive optical decoders had different numbers of trainable diffractive layers in each model. Figure 5 reports the output results of three exemplary handwritten digits and a test grating object information transferred through a new unknown random diffuser using these three hybrid models with $K = 2$, 4, and 6. The visualization of the decoded images and the corresponding PCC values in Fig. 5 illustrate that the 2-layer diffractive decoder network had a relatively lower image reconstruction performance through an unknown random diffuser (despite still providing recognizable images at the output plane). Our results further reveal that the information transmission quality improved significantly as we increased the number of diffractive layers in the decoder architecture, as shown in Fig. 5. The fact that the $K = 2$ decoder still worked (despite a compromise in its output image quality) indicates the effectiveness of the joint training strategy, where the electronic encoder CNN can efficiently collaborate with the diffractive optical decoder and partially compensate for its reduced degrees of freedom caused by fewer diffractive layers in its architecture.

*2.3 Influence of limited phase bit depth*

In the analyses presented above, we did not impose a bit depth restriction during the training and testing stages, and the encoded phase patterns and the corresponding diffractive heights on each layer were set with 16-bit. In certain applications, the available bit depth of the encoded phase pattern would be limited due to e.g., the resolution of the spatial light modulators (SLM); similarly, the bit depth of the diffractive features will be constrained by the 3D fabrication resolution. Motivated to mitigate these challenges, we quantized the electronic encoder and the diffractive decoder that were trained with 16-bits into lower quantization levels in the test stage and investigated the influence of the limited bit depths of the encoded phases patterns and the diffractive layers on the quality of the optical information transfer through random new diffusers. Fig. 6(a) reports the results of this comparative analysis for several combinations of encoded phase patterns and diffractive decoders with different bit depths. The unconstrained electronic-optical hybrid model's output results are shown at the top for reference. In this comparison, we first kept the bit depth of the encoded phase patterns as 16-bit while quantizing the diffractive decoder trained with 16-bit to 2-bit, 3-bit, or 4-bit – performing a form of ablation study. The output images of three MNIST handwritten digits and a grating-like object are displayed in Fig. 6(a), revealing a quick improvement in the information transfer performance with increasing diffractive decoder bit depth. We utilized the diffractive decoder (trained under 16-bit depth) using merely 2-bits, and the output image was blurry; however, the output of the 3-bit diffractive decoder improved significantly and the gratings were resolvable (see Fig. 6(a)). As we further increased the bit depth to 4, the output performance became approximately identical to that of the same model without any bit-depth constraints, as shown in Fig. 6(a).

On the other hand, when we limited the bit depth of the encoded phase patterns while maintaining the diffractive decoder at 16 bits, it was surprising to find out that even with only 1-bit, i.e., *binary encoded phase patterns*, the encoder-decoder system remained capable of transferring the input images of interest through unknown random phase diffusers with acceptable output performance (Fig. 6(a)). Furthermore, the results obtained using the 2-bit phase encoder were comparable to the unconstrained model (Fig. 6(a) top). These ablation studies highlight the strong robustness of our information encoding strategy to the changes in the phase bit depth that is available. Figure 6(b) reports additional results for various combinations of quantized encoded phase and quantized diffractive decoder, showing that a larger phase bit depth generally leads to better output image performance.

These earlier results were ablation studies performed after the joint training of the electronic-optical model, by reducing the available bit depth at the testing phase of the encoder-decoder pair. To explore the minimum requirement of bit depth for accurately transferring optical information through random diffusers, next we adopted the bit-depth limitation during the training process and created three additional electronic-optical models with different levels of bit depth available; see Fig. 7. When both the encoded phase patterns and the diffractive optical decoder were constrained to only 2 bits, the transferred information suffered from significant distortions at the output; however, the contours of some images were still visible. Keeping a bit depth of 2 for the phase encoder while increasing the diffractive decoder's bit-depth to 4 achieved much better information transfer, where the spatial details of the input images were successfully reconstructed after passing through random phase diffusers. For the combination where both the electronic phase encoder and the optical diffractive models had 4 bits of phase, the performance of optical information transfer through random diffusers was comparable to the hybrid model trained and tested with 16 bits (see Fig. 7).

*2.4 Experimental demonstration of optical information transfer through unknown random phase diffusers using a diffractive decoder with electronic encoding*

The presented hybrid electronic-optical model was demonstrated experimentally based on a terahertz continuous wave system, operating at $\lambda = 0.75$mm, as shown in Fig. 8(c-d) (see the Materials and methods section). The resulting phase patterns encoded from the input images by the electronic encoder CNN were physically 3D-printed and illuminated by the terahertz source, propagating forward through a 3D-printed unknown random diffuser and the diffractive decoder layers, positioned sequentially. For this proof-of-concept experimental demonstration, we jointly trained an electronic encoder CNN and a diffractive decoder with two layers ($K=2$), each consisting of 120×120 diffractive neurons with a size of 0.4mm. Owing to the limitations posed by the 3D printer's resolution and build size, we adjusted the dimensions of the optical elements and operated at a Fresnel number of 9.4, thereby maintaining the difficulty of the information transfer task through random diffusers (see Fig. 2(c)). This electronic-optical model was trained with $n=20$ different random diffusers for 100 epochs, i.e., $N=2000$ different random diffusers were utilized during the training, each with a correlation length of $L \sim 5\lambda$. Moreover, an additional output power efficiency loss was introduced to make the diffractive decoder strike a good balance between the information transfer performance and diffraction efficiency (see the Materials and methods section). To mitigate the negative impact of potential misalignments during the fabrication and assembly of the optical components, we also introduced random displacements to the diffractive layers (on purpose) in the training process, "vaccinating" the diffractive decoder against such imperfections (see Methods section for details)[38]. Figure 9 shows the experimental results of the optical information transfer through unknown random phase diffusers using a 3D-printed diffractive decoder and electronic encoding. The experimental measurements match the simulation results very well, confirming the feasibility of the hybrid electronic-optical model to effectively transfer optical information and messages of interest through random unknown phase diffusers.

## 3  Discussion

We presented a joint electronic encoder and optical decoder model designed to transfer optical information through random unknown phase diffusers, outperforming (1) an ideal (diffraction limited) imaging system and (2) a system solely employing trainable diffractive surfaces as demonstrated in our previous work[30]. With an electronic encoder CNN encoding the original input images into 2- to 4-bit depth phase patterns, a jointly-trained diffractive optical decoder becomes much more resilient to the distortions caused by random, unknown phase diffusers along the optical path, leading to enhanced performance and higher data transmission fidelity. The diffractive optical decoder, consisting only of passive diffractive layers, can decode the encoded information through random phase diffusers at the speed of light and enable the hybrid model to work with low power consumption. The overall volume of the diffractive optical decoder is also compact, with an axial span of < 70$\lambda$. Our experimental results in terahertz part of the spectrum confirmed the applicability of our hybrid electronic-encoder and optical-decoder model for transferring optical information through unknown random diffusers.

Following the previous literature on random phase diffusers[21,22,39,40], in this work we used thin optical elements with random phase patterns to simulate diffusers; the exploration of volumetric diffusers, which can be described by several thin phase diffusers and split-step beam propagation, is an exciting future research direction[41,42]. Additionally, wavelength-division multiplexing, widely used in fiber-optic communication to enable high transmission bandwidths[43], can be integrated into the presented jointly trained models, allowing simultaneous information transfer at multiple wavelengths and increasing the overall capacity of information transfer through random unknown diffusers. Finally, our method can be physically scaled (expanded/shrunk) with respect to the illumination wavelength without retraining its components and can operate at different parts of the electromagnetic spectrum to transfer optical information through random scattering media.

## 4   Materials and methods

*4.1 Electronic encoder design*

We used a convolutional neural network (CNN) to encode the input object into a 2D phase pattern. The network architecture is shown in Supplementary Fig. S1, which has convolutional layers with 3 × 3 filters followed by Batch Normalization (BN)[44] and Parametric Rectified Linear Unit (PReLU)[45]. We performed downsampling directly by max pooling. The network ends with two fully connected layers and a sigmoid activation layer.

*4.2 All-optical diffractive decoder model*

The monochrome illumination with a wavelength of $\lambda$ was used for optical information transfer through unknown random diffusers. The diffractive layers were modeled as thin optical modulation elements where the complex transmission coefficient of the layer $m$ located at $z = z_m$ can be modeled as:

$$t_m(x, y, z_m) = a_m(x, y, z_m) \exp(j\phi_m(x, y, z_m)) \qquad (1)$$

Both the amplitude modulation $a_m(x, y, z_m)$ and the phase modulation $\phi_m(x, y, z_m)$ can be written as a function of the diffractive neuron's thickness $h_m(x, y, z_m)$ and the incident wavelength $\lambda$:

$$a_m(x, y, z_m) = \exp\left(-\kappa(\lambda)\frac{2\pi h_m(x, y, z_m)}{\lambda}\right) \qquad (2)$$

$$\phi_m(x,y,z_m) = (n(\lambda) - n_{air})\frac{2\pi h_m(x,y,z_m)}{\lambda} \tag{3}$$

where $n(\lambda)$ and $\kappa(\lambda)$ are the refractive index and the extinction coefficient of the material, i.e., the complex refractive index can be written as $\tilde{n}(\lambda) = n(\lambda) + j\kappa(\lambda)$. In this work, the height of each neuron was defined as:

$$h_m(x,y,z_m) = \frac{h_{max}}{2} \cdot \left(\sin\left(h_p(x,y,z_m)\right) + 1\right) + h_{base} \tag{4}$$

where $h_p$ is the variable optimized during the data-driven training procedure. The actual height of each diffractive neuron $h_m(x,y,z_m)$ was calculated by setting $h_{max} = 1.33\lambda$ and a fixed base thickness of $h_{base} = 0.67\lambda$.

We modeled the random diffusers as pure phase elements, whose complex transmission $t_D(x,y)$ can be calculated using the refractive index difference between the air and the diffuser material ($\Delta n = n(\lambda) - n_{air} \approx 0.725$). The random height map $h_D(x,y)$ was defined as:

$$h_D = rem\left(W * K(\sigma) + h_{base}, \frac{\lambda_{max}}{n(\lambda) - n_{air}}\right) \tag{5}$$

where $rem(.)$ denotes the remainder after division. $W$ is a random height matrix on which each pixel, i.e., $W(x,y)$, and it follows a normal distribution with a mean of $\mu$ and a standard deviation of $\sigma_0$, i.e.

$$W(x,y) \sim \mathcal{N}(\mu, \sigma_0) \tag{6}$$

$K(\sigma)$ is a zero-mean Gaussian smoothing kernel with a standard deviation of $\sigma$, and '$*$' denotes the 2D convolution operation. The phase-autocorrelation function $R_d(x,y)$ of a random phase diffuser ($t_D(x,y)$) is calculated as:

$$R_d(x,y) = \exp(-\pi(x^2 + y^2)/L^2) \tag{7}$$

where $L$ refers to the correlation length of the random diffuser. By numerically fitting the function $\exp(-\pi(x^2 + y^2)/L^2)$ to $R_d(x,y)$, we can statistically get the correlation length $L$ of randomly generated diffusers. In this work, for $\mu = 25\lambda$, $\sigma_0 = 8\lambda$ and $\sigma = 4\lambda$, we calculated the average correlation length as $L \sim 5.3\lambda$ based on 2,000 randomly generated phase diffusers.

Adjacent optical elements (e.g., input phase patterns, random diffusers, diffractive layers, and detectors) are optically connected by free space light propagation in air, which was formulated using the Rayleigh-Sommerfeld equation[46]. The propagation can be modeled as a shift-invariant linear system with the impulse response:

$$w(x, y, z) = \frac{z}{r^2}\left(\frac{1}{2\pi r^2} + \frac{1}{j\lambda}\right)\exp\left(\frac{j2\pi r}{\lambda}\right) \quad (8)$$

where $r = \sqrt{x^2 + y^2 + z^2}$ and $j = \sqrt{-1}$.

Considering a plane wave that is incident at a phase-modulated object $o(x, y, z = 0)$ positioned at $z = 0$, we can write the distorted image right after the random phase diffuser $t_D(x, y)$ located at $z_0$ as:

$$u_0(x, y, z_0) = t_D(x, y) \cdot [o(x, y, z = 0) * w(x, y, z_0)] \quad (9)$$

This distorted field is used as the input field of subsequent diffractive layers. And the optical field $u_m(x, y, z_m)$ right after the $m^{th}$ diffractive layer at $z = z_m$ can be written as:

$$u_m(x, y, z_m) = t_m(x, y) \cdot [u_{m-1}(x, y, z_{m-1}) * w(x, y, \Delta z_m)] \quad (10)$$

where $\Delta z_m = z_m - z_{m-1}$ is the axial distance between two successive diffractive layers. After being modulated by all the $M$ diffractive layers, the optical field was collected at an output plane which was $\Delta z_d = 13.3\lambda$ away from the last diffractive layer. The intensity is used as the decoded output intensity profile of the diffractive decoder:

$$I(x, y) = |u_M(x, y, z_M) * w(x, y, \Delta z_d)|^2 \quad (11)$$

*4.3 Quantization of the encoded phase and the diffractive decoder layers*

To evaluate the performance of the hybrid electronic-encoder and optical-decoder models under limited bit-depth cases, we quantized the encoded phase and/or the heights of the features on the diffractive layers into lower quantization levels. For the encoded phase patterns with a maximum phase of $\pi$, the range $[0, \pi]$ was equally divided into $2^{bit\ depth}$ steps. For example, the 1-bit encoded phase is binary encoded with 0 and $\pi$, while the 2-bit encoded case has $\left\{0, \frac{\pi}{3}, \frac{2\pi}{3}, \pi\right\}$, i.e., four steps. In the same way, we quantized the diffractive decoder heights in the range of $[h_{base}, h_{base} + h_{max}]$.

*4.4 Terahertz experimental setup and design*

The experimental setup is illustrated in Fig. 8(b). The terahertz source used in the experiment was a modular amplifier (WR9.0M SGX, Virginia Diode Inc.)/multiplier chain (WR4.3x2 WR2.2x2, Virginia Diode Inc.) (AMC) with a compatible diagonal horn antenna (WR2.2, Virginia Diode Inc.). The input of AMC was a 10dBm RF input signal at 11.1111 GHz ($f_{RF1}$) and after being multiplied 36 times, the output signal was continuous-wave (CW) radiation at 0.4THz. The AMC was also modulated with a 1kHz square wave for lock-in detection. The distance between the exit aperture of the horn antenna and the object plane of the 3D-printed diffractive optical network was about 75cm. After passing the diffractive decoder network, the output signal was 2D scanned with a 1.2mm step by a single-pixel Mixer (WRI 2.2, Virginia Diode Inc.) placed on an XY positioning stage, built by combining two linear motorized stages (NRT100, Thorlabs). A 10 dBm RF signal at 11.0833 GHz ($f_{RF2}$) was sent to the detector as a local oscillator to down-convert the signal to 1 GHz for further measurement. The down-converted signal was amplified by a low-noise amplifier (ZRL-1150-LN+, Mini-Circuits) and filtered by a 1 GHz (+/-10 MHz) bandpass filter (Electronics 3C40-1000/T10-O/O, KL). The signal first passed through a low-noise power detector (ZX47-60, Mini-Circuits) and then was measured by a lock-in amplifier (SR830, Stanford Research) with the 1kHz square wave used as the reference signal. The lock-in amplifier readings were calibrated into a linear scale.

The phase objects, diffusers, and diffractive layers used for experimental demonstration were fabricated using a 3D printer (Objet30 Pro, Stratasys). The absorption of the 3D printing material was taken into account and estimated by Eq. (2) for the phase patterns, diffusers and diffractive layers during the training. For the two-layer decoder model used in the experimental demonstration, each diffractive layer consisted of 120×120 diffractive neurons, each with a lateral size of 0.4mm. The diffuser had the same size as the diffractive layers with 48×48 mm$^2$ (120×120 pixels). The size of the input and output FOV were designed as 33.6×33.6 mm$^2$ (84×84 pixels). The axial distances between the input phase object and the random diffuser, the diffuser and first diffractive layer, two successive diffractive layers, and the last diffractive layer and the output plane were set to be 53.3$\lambda$, 40$\lambda$, 40$\lambda$ and 26.7$\lambda$, respectively.

To build the resilience of the diffractive decoder layers to potential mechanical misalignments in the experimental testing, the diffractive decoder was "vaccinated" with deliberate random shifts during the training[38]. For this vaccination process, a random lateral displacement $(D_x, D_y)$ was added to the diffractive layers, where $D_x$ and $D_y$ were randomly and independently sampled, i.e.,

$$D_x \sim \mathbf{U}[-0.5\lambda, 0.5\lambda], D_y \sim \mathbf{U}[-0.5\lambda, 0.5\lambda] \qquad (12)$$

A random axial displacement of $D_z$ was also added to the axial separations between any two consecutive diffractive planes. Accordingly, the axial distance between any two consecutive diffractive layers was set to $d + D_z$, where $d$ is the designed distance and $D_z$ was randomly sampled,

$$D_z \sim \mathbf{U}[-1.5\lambda, 1.5\lambda] \qquad (13)$$

Considering the material absorption and minimum detectable signal in our experiments, we also added another power penalty to balance the output quality and the diffraction efficiency. We calculated the power efficiency $E(I_{output})$ of the diffractive network as:

$$E(I_{output}) = \frac{\sum I_{output}(x,y)}{\sum |o(x,y)|^2} \qquad (14)$$

and the corresponding diffraction efficiency penalty was calculated as follows:

$$Loss_{eff} = \max\{0, E_{target} - E(I_{output})\} \tag{15}$$

where $E_{target}$ was the target power efficiency, which was set to 0.04 empirically.

*4.5 Image contrast enhancement*

To better visualize the images, we digitally enhanced the contrast of the experimental measurements using a built-in MATLAB function (*imadjust*), which by default saturates the top 1% and the bottom 1% of the pixel values and maps the resulting images to a dynamic range between 0 and 1. All the quantitative data analyses, including PCC calculations and resolution test target results, are based on raw data without applying image contrast enhancement.

*4.6 Simulation of the standard lens-based image transmission system*

We numerically implemented a conventional lens-based image transmission system to evaluate the impact of a given random diffuser on the output image. A Fresnel lens was designed to have a focal length ($f$) of $136.1\lambda$ and a pupil diameter of $96\lambda$, whose transmission coefficient $t_L(x, y)$ was formulated as:

$$t_L(x, y) = P(x, y) \exp\left(-j\frac{\pi}{\lambda f}(x^2 + y^2)\right) \tag{16}$$

where $P(x, y)$ is the pupil function:

$$P(x, y) = \begin{cases} 1, & \sqrt{x^2 + y^2} < 96\lambda \\ 0, & otherwise \end{cases} \tag{17}$$

The lens was placed $2f$ ($272.2\lambda$) away from the input objects and the image plane was also $2f$ ($272.2\lambda$) away from the lens. The light propagation was calculated using the angular spectrum method. And the intensity profiles at the image plane were collected as the results of imaging through random phase diffusers using an ideal diffraction-limited lens.

## 5  Figures and Figure Captions

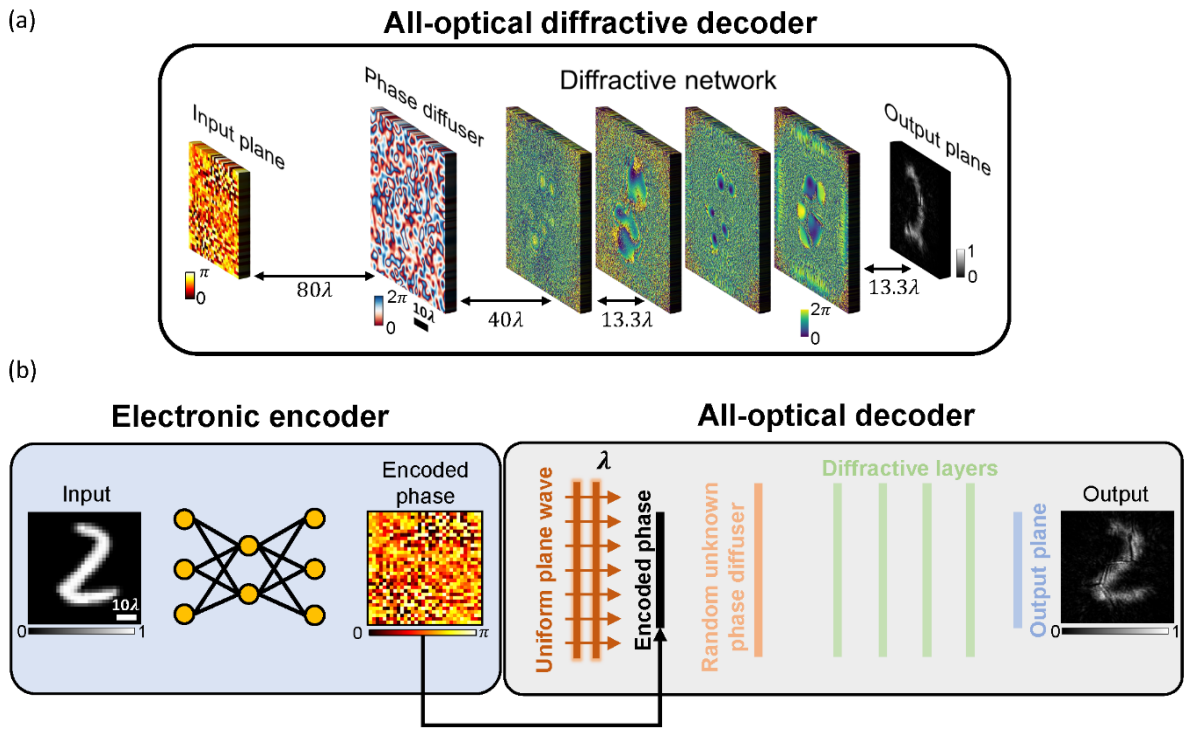

**Fig. 1** The pipeline of the hybrid electronic encoder and the optical diffractive decoder for optical information transfer through random unknown diffusers. (a) The schematic drawing of the presented diffractive decoder, which all-optically decodes the encoded information distorted by random phase diffusers without the need for a digital computer. (b) The workflow of the hybrid electronic-optical model: the electronic neural network encodes the input objects into 2D phase patterns and the all-optical diffractive neural network decodes the information transmitted through random, unknown phase diffusers.

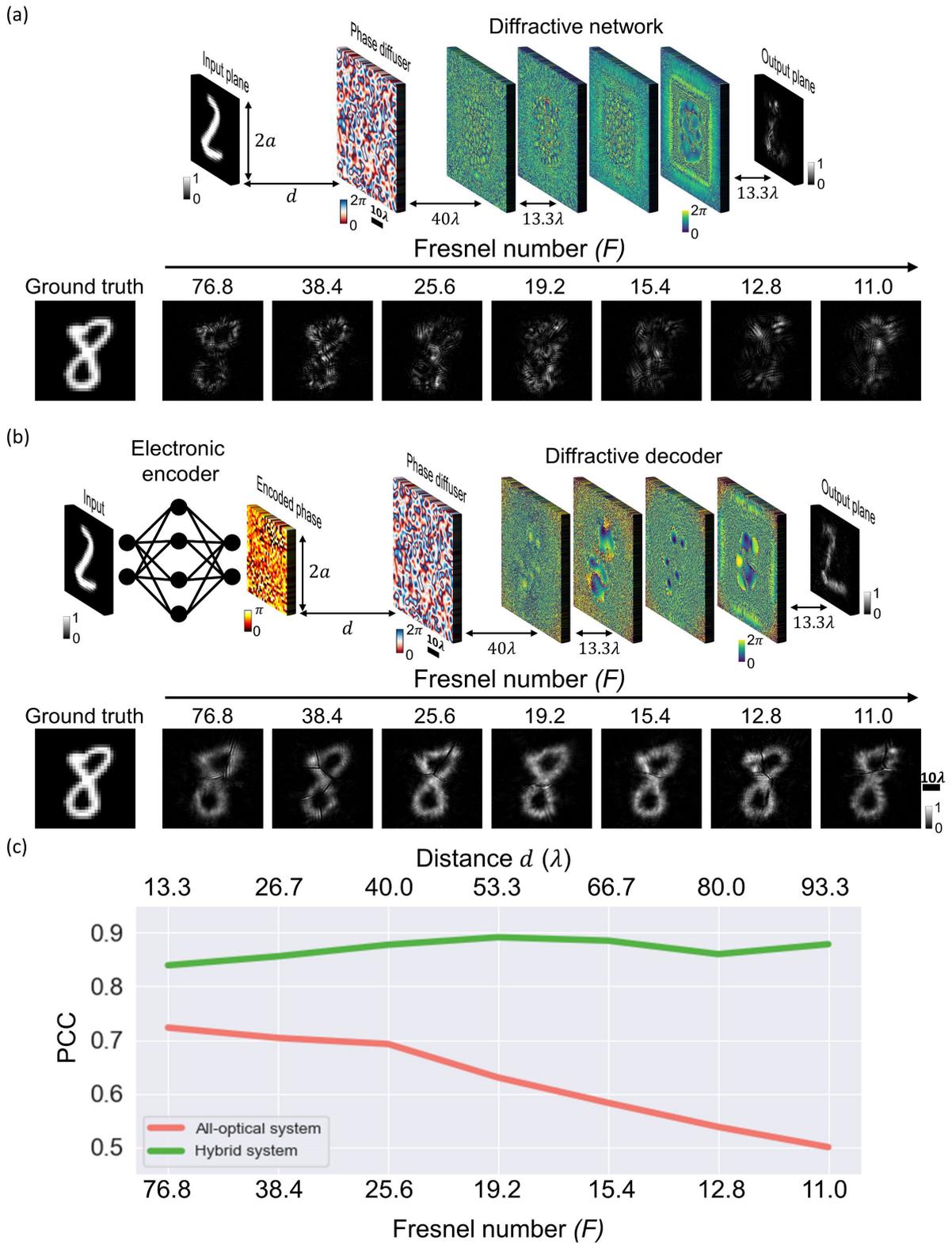

**Fig. 2** Comparison between all-optical diffractive networks and the hybrid electronic-optical models for transferring optical information through random unknown diffusers (a) Schematic of a 4-layer diffractive network trained to all-optical reconstruct the amplitude images of input objects through random phase diffusers without electronic encoding. (b) Schematic of a 4-layer diffractive decoder with an electronic encoder jointly trained to decode the encoded optical image through random unknown diffusers (c) The information transmission fidelity (PCC) of the two approaches (all-optical vs. hybrid) as a function of the Fresnel number ($F$) of the diffractive

system. All the parameters were identical in the two approaches being compared, except for the existence of the electronic encoder (at the front-end of the hybrid approach).

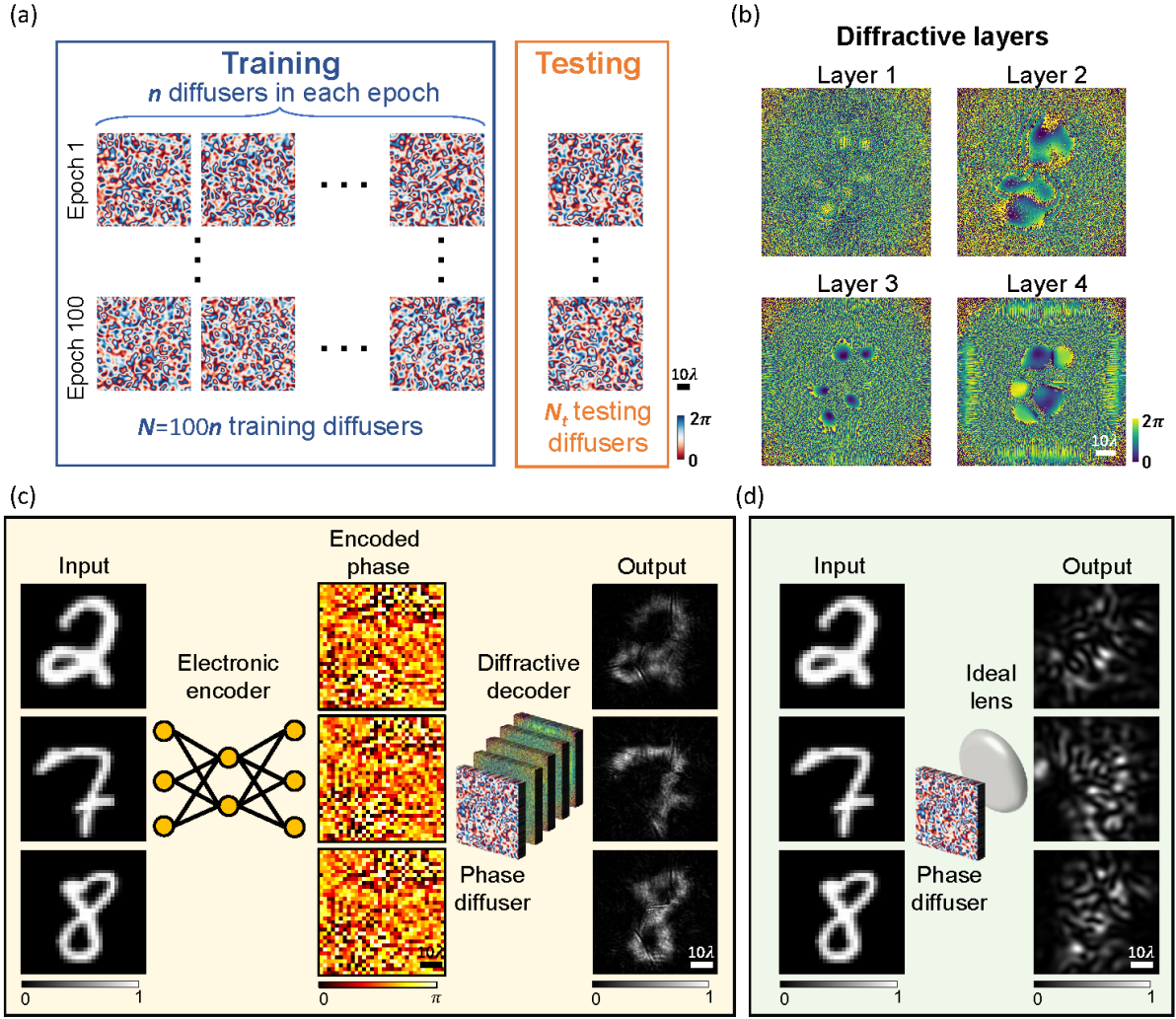

**Fig. 3** Optical information transfer through random phase diffusers using electronic encoding and diffractive decoding. (a) Random phase diffusers with a correlation length $L = 5\lambda$ used for training and testing the presented joint electronic-optical model. (b) The phase profiles of the trained diffractive layers of the optical decoder. (c) Examples of optical information encoded by an electronic encoder CNN that is jointly trained with an optical decoder; output images processed by the all-optical decoder are also shown on the right. (d) Sample images captured by an aberration-free diffraction-limited lens.

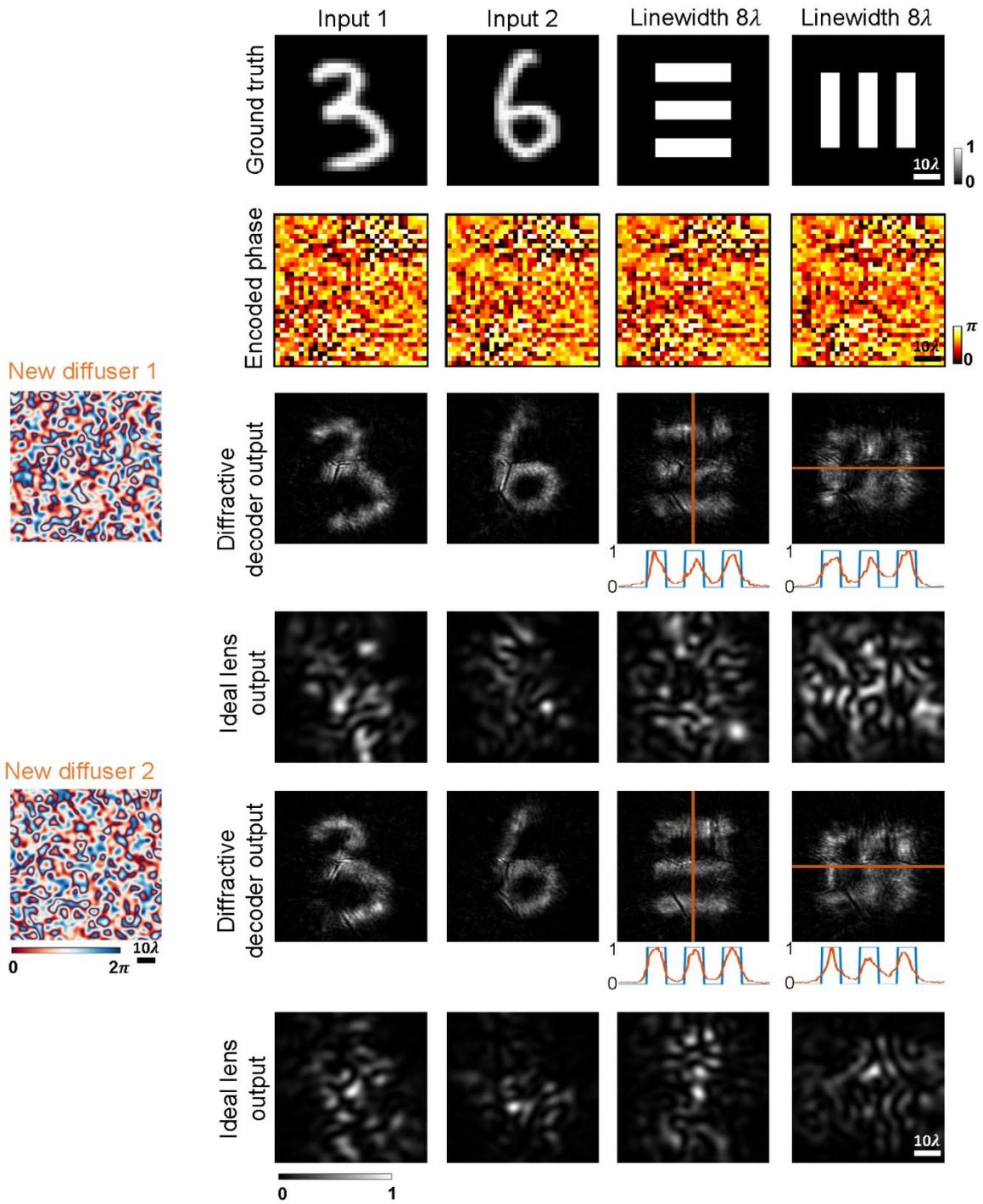

**Fig. 4** Simulation results of the hybrid electronic-optical model for optical information transmission through unknown random phase diffusers. $L = 5\lambda$.

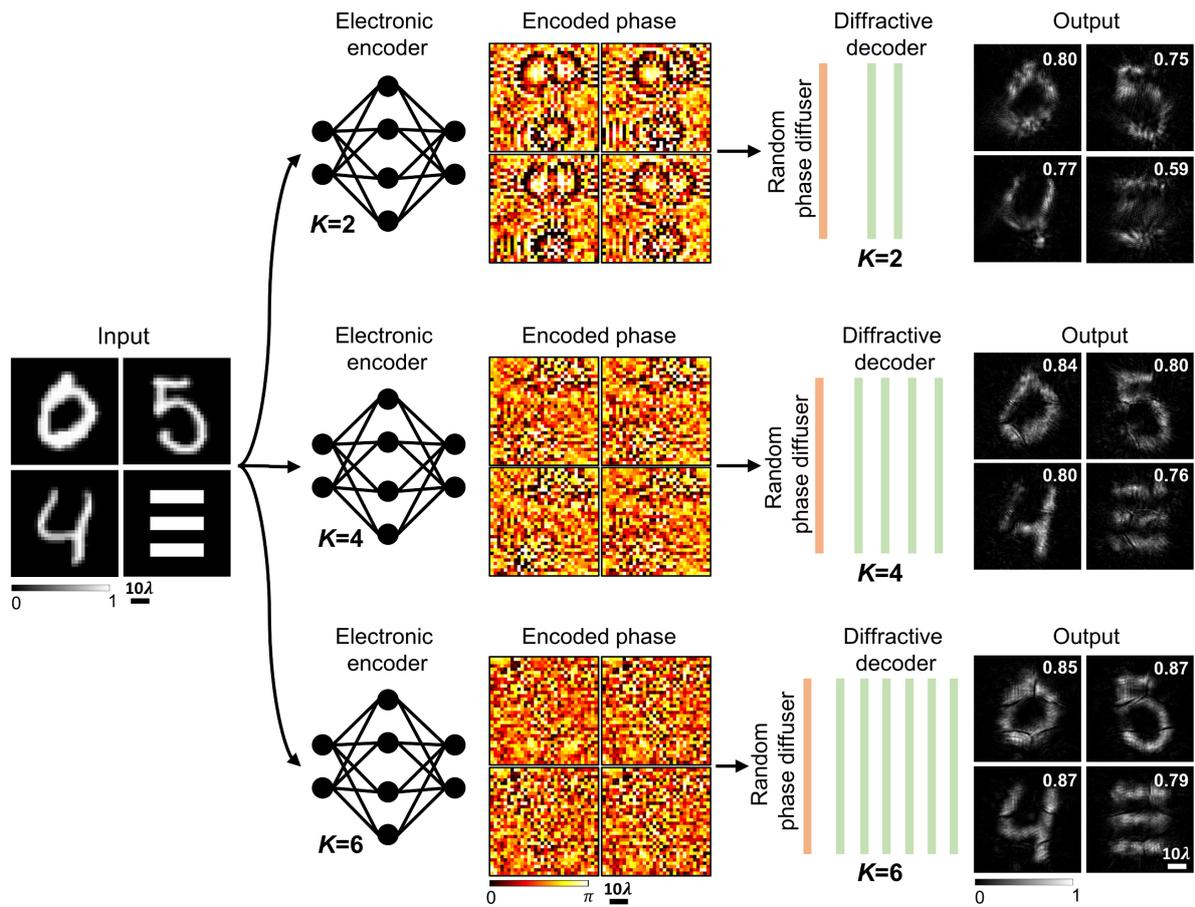

**Fig. 5** Additional trainable diffractive layers (increasing *K*) improve the optical information transfer performance through random unknown diffusers. The PCC values are listed for each transmitted message.

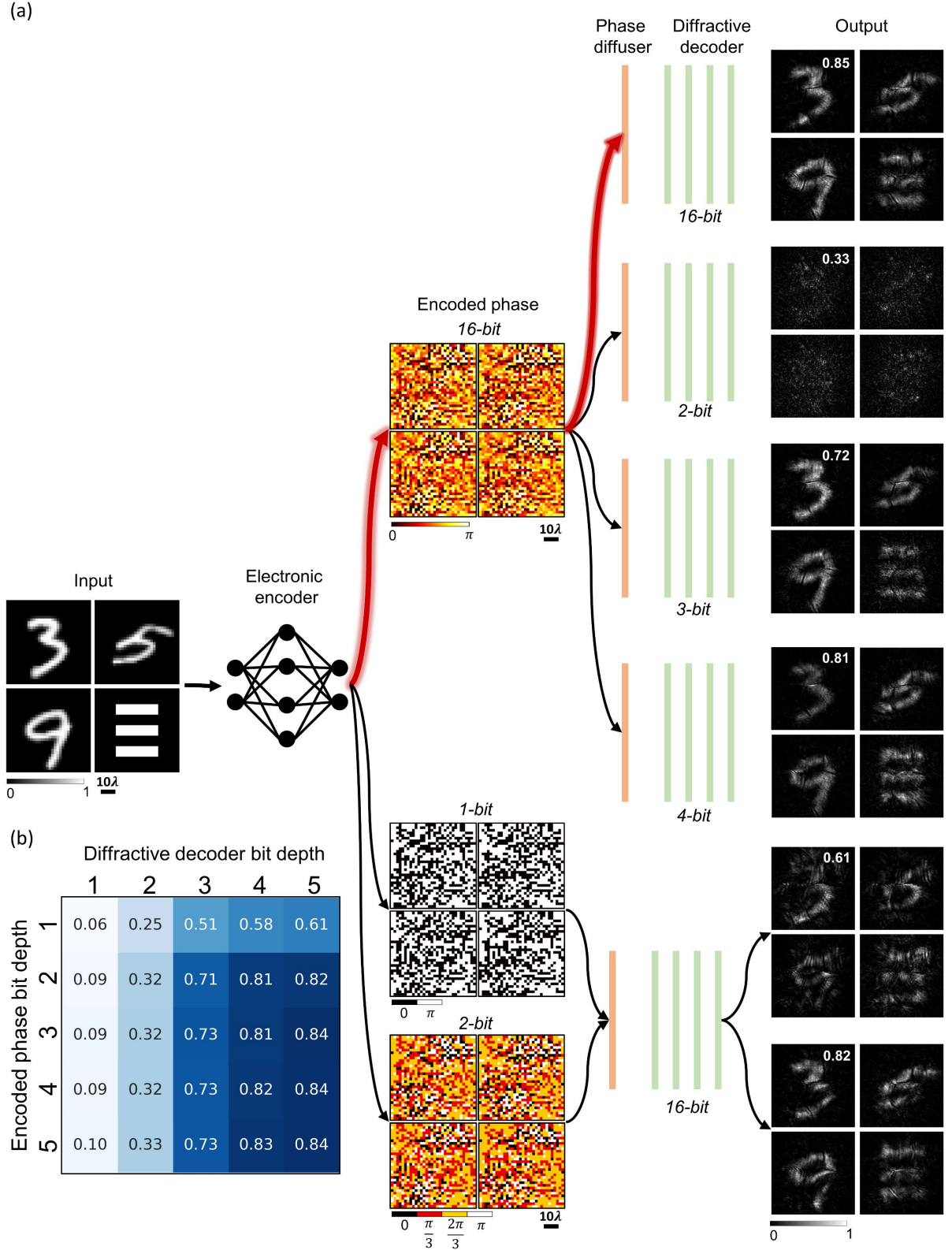

**Fig. 6** The impact of the phase bit depth on the optical information transfer fidelity. (a) The joint model trained without any bit-depth limitations, i.e., 16-bit phase representation, and tested with restricted bit-depth phase patterns and decoder pairs. The average PCC value for 10,000 handwritten test digits transmitted through $n = 20$ unknown (new) random diffusers is provided in each case. (b) The resulting mean PCC values are compared for various combinations of phase bit-depth. $L = 5\lambda$.

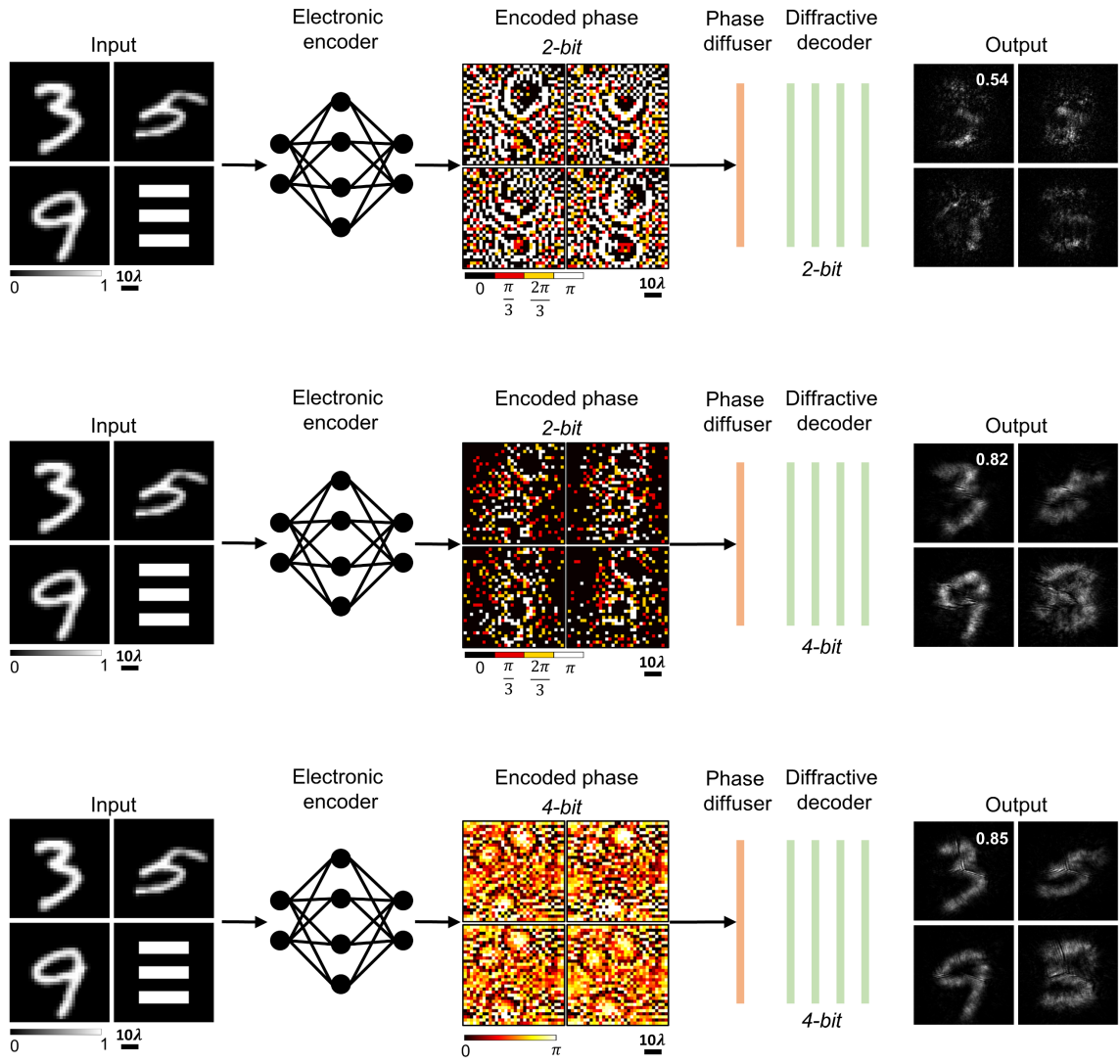

**Fig. 7** The electronic encoder and the diffractive decoder trained and tested with different levels of phase bit-depth. The average PCC values are listed for each case. $L = 5\lambda$.

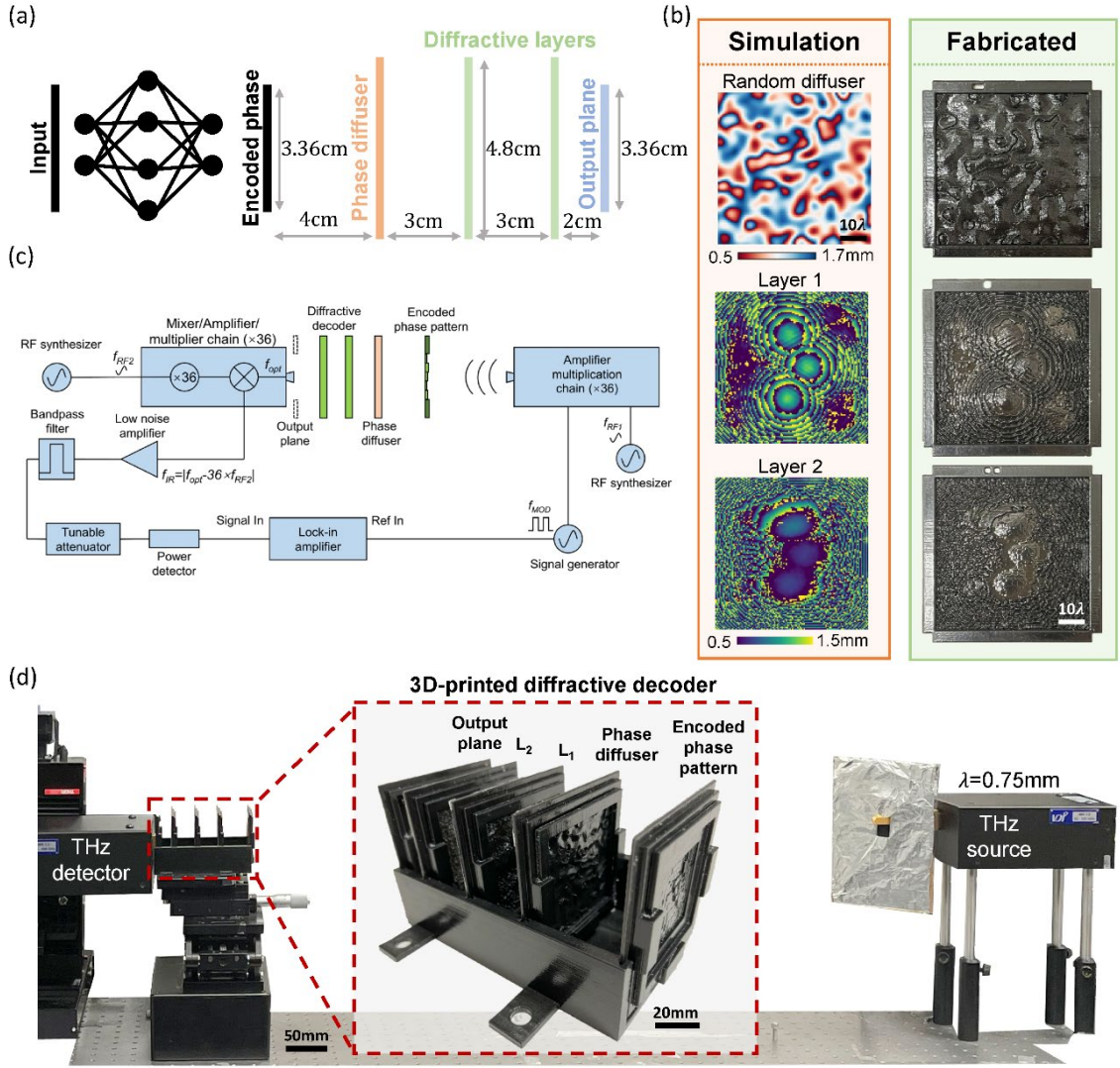

**Fig. 8** Experimental demonstration of optical information transfer through an unknown random diffuser using a jointly trained pair of an electronic encoder and an optical decoder. (a) Schematic of the joint model for experimental demonstration. (b) Left: Height profiles of a new random diffuser (never seen before) and the trained diffractive layers of the all-optical decoder. Right: The fabricated new random diffuser and the diffractive layers used in the experiment. (c) Schematic of the terahertz system. (d) Photograph of the experimental setup and the 3D-printed diffractive decoder.

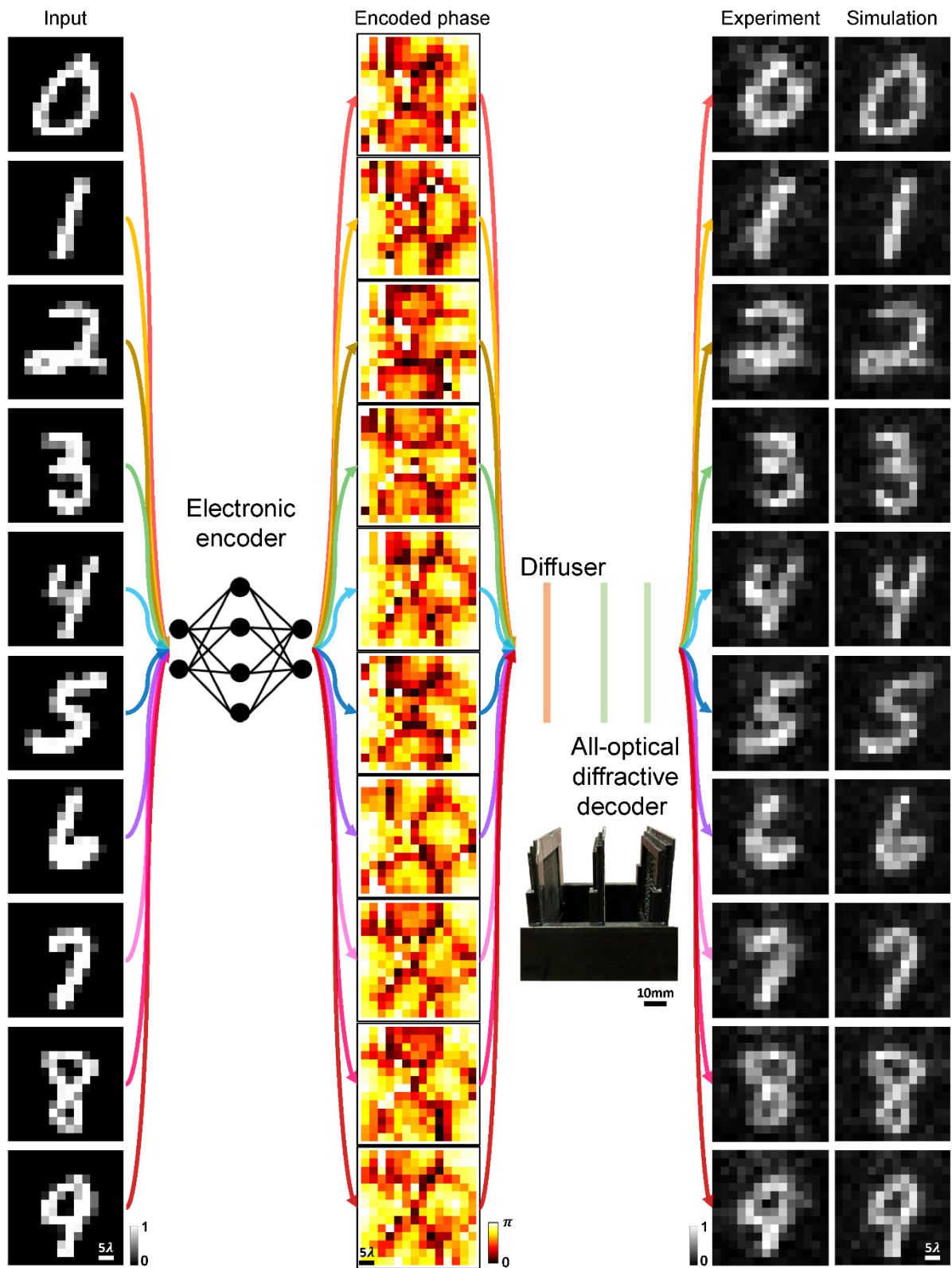

**Fig. 9** Experimental results of optical information transfer through an unknown random phase diffuser using the 3D-printed diffractive decoder with electronic encoding.

# 6   Appendix

Appendix includes:
**Digital Implementation Details**
**Fig. S1** The architecture of the electronic encoder neural network (CNN).
**Video 1** Results of the electronic encoder and the diffractive decoder (jointly-trained) that transfer different optical images of interest through random unknown phase diffusers which are constantly evolving. (MP4, 13.3 MB)